\documentclass[
reprint,
superscriptaddress,
preprintnumbers,
nobibnotes,nofootinbib,
amsmath,amssymb,
aps,
floatfix,
]{revtex4-1}

\usepackage[normalem]{ulem}
\usepackage{gensymb}
\usepackage{upgreek}
\usepackage{graphicx}
\usepackage{dcolumn}
\usepackage{bm}
\usepackage{floatrow}
\usepackage{float}
\usepackage{hyperref}
\usepackage{xcolor}
\usepackage{textcomp}
\usepackage{amsmath}
\usepackage{varwidth}
\usepackage{gensymb}
\usepackage{siunitx}
\hypersetup{
    colorlinks,
    linkcolor={red!50!black},
    citecolor={blue!50!black},
    urlcolor={blue!80!black}
}
\usepackage[center]{titlesec}
  \titleformat{\section}
   {\centering\normalfont\fontsize{10}{10}\bfseries}{\thesection}{1em}{}
  \titlespacing{\section}{0pt}{12pt plus 4pt minus 2pt}{6pt plus 2pt minus 2pt}
    \titleformat{\subsection}
   {\centering\normalfont\fontsize{10}{10}\bfseries}{\thesubsection}{1em}{}
  \titlespacing{\subsection}{0pt}{12pt plus 4pt minus 2pt}{6pt plus 2pt minus 2pt}
  
\newcommand{\unm}{Center for High Technology Materials and Department of Physics and Astronomy, University of New Mexico, Albuquerque, NM, USA}

\begin{document}

\title{Stimulated emission depletion microscopy with diamond silicon-vacancy centers} 
\date{\today}

\author{Yaser Silani$^{\mathsection}$}
\affiliation{\unm}
\author{Forrest Hubert$^{\mathsection}$}
\affiliation{\unm}
\author{Victor Acosta}
\email{vmacosta@unm.edu}
\affiliation{\unm} 
\renewcommand{\thefootnote}{}{\footnote{$\mathsection$ These authors contributed equally to this work.}}
\date{\today} 

\begin{abstract}
The spatial resolution and fluorescence signal amplitude in stimulated emission depletion (STED) microscopy is limited by the photostability of available fluorophores. Here, we show that negatively-charged silicon vacancy (SiV) centers in diamond are promising fluorophores for STED microscopy, owing to their photostable, near-infrared emission and favorable photophysical properties. A home-built pulsed STED microscope was used to image shallow implanted SiV centers in bulk diamond at room temperature. The SiV stimulated emission cross section for $765\mbox{--}800~{\rm nm}$ light is found to be $(4.0\pm0.3){\times}10^{-17}~{\rm cm^2}$, which is approximately $2\mbox{--}4$ times larger than that of the negatively-charged diamond nitrogen vacancy center and approaches that of commonly-used organic dye molecules. We performed STED microscopy on isolated SiV centers and observed a lateral full-width-at-half-maximum spot size of $89\pm2~{\rm nm}$, limited by the low available STED laser pulse energy ($0.4~{\rm nJ}$). For a pulse energy of $5~\rm{nJ}$, the resolution is expected to be ${\sim}20~{\rm nm}$. We show that the present microscope can resolve SiV centers separated by $\lesssim150~{\rm nm}$ that cannot be resolved by confocal microscopy. 
\end{abstract}

\maketitle

\section{Introduction}
Stimulated emission depletion (STED) microscopy is one of several techniques which can image fluorescent molecules with a spatial resolution superior to the optical diffraction limit \cite{HEL1994,KLA1999}. While the resolution in STED microscopy can theoretically approach the scale of individual atoms \cite{WES2005}, resolving structures at the few nanometer scale in biological samples remains an experimental challenge. This is partly due to a lack of fluorescent probes which possess the requisite photophysical properties and are sufficiently small, bright, photostable, and non-toxic. 

In STED microscopy, the theoretical lateral resolution, $\Delta d$, scales approximately as $\Delta d\propto \sqrt{I_{\rm sat}/I}$, where $I$ is the optical intensity used to stimulate emission and $I_{\rm sat}$ is the fluorophore's stimulated-emission saturation intensity \cite{HEL2003}. This scaling has two consequences for probe design. The first is that a low $I_{\rm sat}$ is desirable so that low enough values of $I$ can be used to avoid sample photodamage while maintaining high resolution. The second consequence is that a high degree of photostability is required to simultaneously realize low values of $\Delta d$ and a high fluorescence signal amplitude. This is because, when $I_{\rm sat}/I$ is small (as needed for high resolution), many fluorophore absorption events do not produce detectable fluorescence, yet they often have the same propensity for photobleaching \cite{PAW2006}. Thus, if the fluorophore bleaches after a fixed number of absorption events, there is an unavoidable trade off between spatial resolution and fluorescence signal amplitude. A similar argument holds in pulsed STED microscopy, where the STED beam's pulse fluence is substituted for intensity.

Organic dye molecules are among the most widely used fluorophores in STED microscopy \cite{WUR2012}. They can be functionalized to specifically bind to biological targets \cite{SAM2009} and are relatively non-toxic \cite{CHO2009}. They also can produce high fluorescence rates \cite{DEM2011} and feature sufficiently low values of $I_{\rm sat}$ \cite{BOU2013} to enable imaging of cells with a spatial resolution down to ${\sim}20~{\rm nm}$ \cite{DON2006}. Nevertheless, standard organic fluorophores suffer from photobleaching due to irreversible chemical reactions \cite{EGG1998}, thereby limiting the achievable fluorescence signal amplitude and resolution \cite{ORA2017}. 

Solid-state color centers are an intriguing alternative probe for STED microscopy, as the host crystal prevents some forms of photobleaching \cite{AHA2014}. For example, the negatively-charged nitrogen vacancy (NV) color center in diamond exhibits nearly perfect photostability in nanodiamonds with characteristic dimensions down to ${\sim}10~{\rm nm}$ \cite{TIS2009}. Moreover diamond is a relatively non-toxic host crystal that can be functionalized to bind to intracellular targets \cite{MAN2013}. NV centers in bulk diamond have been used to set record spatial resolutions in STED microscopy, with lateral resolutions as small as $\Delta d=2.4~{\rm nm}$ \cite{WIL2012}. However, NV centers have some limitations in their use in STED microscopy. The fluorescence intensity of a single NV center is more than an order of magnitude weaker than a typical organic fluorophore \cite{FAK2010} under similar conditions. They require high stimulated emission depletion intensities, owing to their relatively low cross section (approximately $1\mbox{--}2\times10^{-17}~{\rm cm^2}$ \cite{HAN2009,RIT2009}) and their propensity for excited state absorption \cite{ASL2013,HAC2018}. Finally, NV centers tend to blink in small nanodiamonds and do not produce observable fluorescence in nanodiamonds smaller than ${\sim}10~{\rm nm}$~\cite{RAB2007,TIS2009}.

\begin{figure*}[htbp]
\centering
\includegraphics[width=\textwidth]{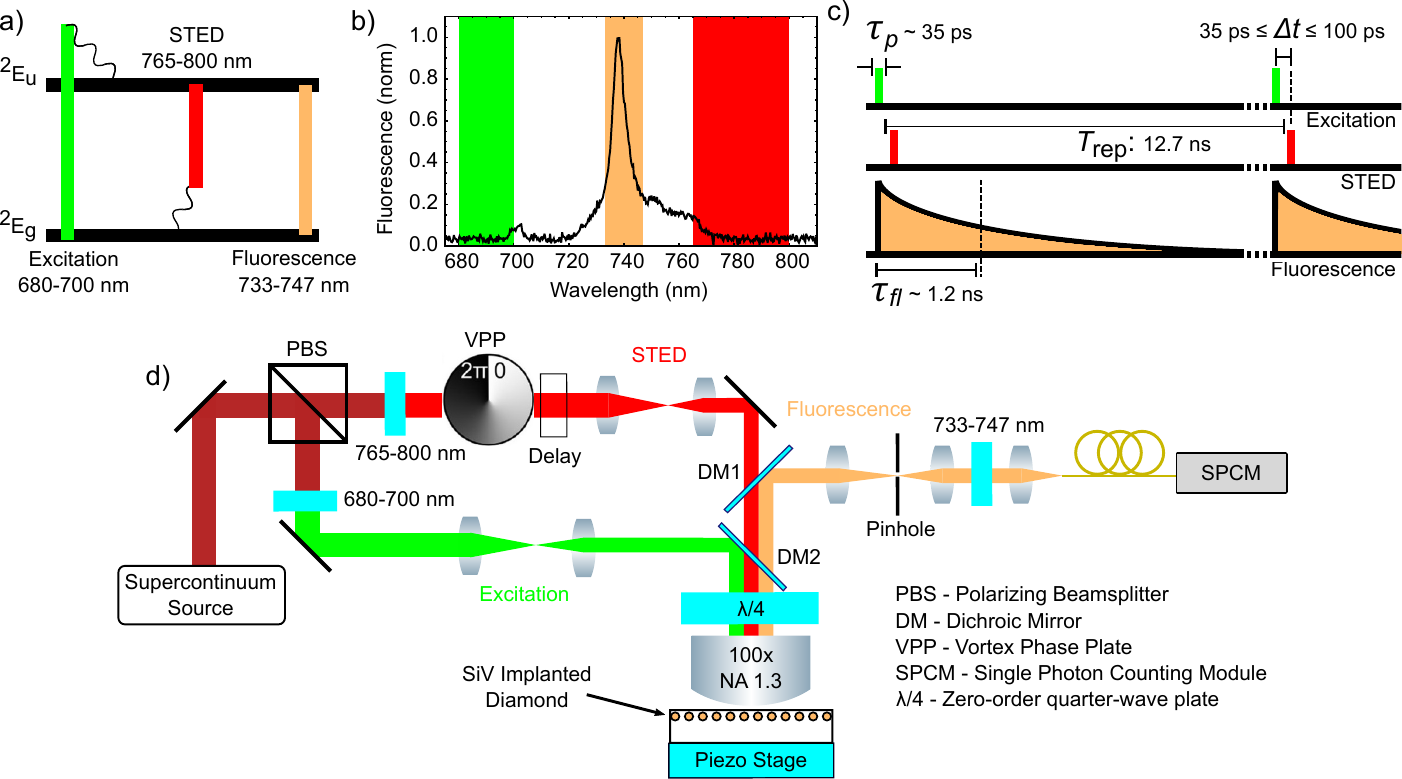}\hfill
\caption{\label{fig:experimental setup}
\textbf{SiV STED methodology.} (a) SiV optical transitions between the ground ($^{2}E_g$) and excited ($^{2}E_u$) electronic states \cite{HEP2014,ROG2014}. (b) SiV fluorescence spectrum under $685\mbox{-}{\rm nm}$ excitation. The bands used for excitation, fluorescence collection, and STED, are labeled in green, orange, and red, respectively. The wavelength band of the STED beam ($765\mbox{--}800~{\rm nm}$) was selected to maximize the available power from the supercontinuum source, while minimizing anti-Stokes excitation (Sec.~\ref{sec:SI Anti-stokes}) and remaining within the tail of the phonon sideband. (c) Pulse sequence used for STED microscopy and for measuring the SiV stimulated emission cross section. The pulses are provided by a picosecond supercontinuum source with a repetition rate of $78~{\rm MHz}$. (d) Apparatus used for STED microscopy of SiV centers. An oil-immersion microscope objective with numerical aperture $\rm NA=1.3$ focuses light onto, and collects fluorescence from, SiV centers. Additional details are found in Sec.~\ref{sec:SI Setup}.}
\end{figure*}

Negatively-charged silicon vacancy (SiV) color centers in diamond may offer a more promising alternative for STED microscopy applications. SiV centers have been shown to be photostable in nanodiamonds as small as ${\sim}2~{\rm nm}$ \cite{VLA2014}, and their fluorescence spectrum lies in a narrow band in the near infrared \cite{WAN2006}. Here, we report measurements of the stimulated emission cross section of SiV centers in bulk diamond. We find $\sigma_{\rm STED}=(4.0\pm0.3){\times}10^{-17}~{\rm cm^2}$ for $765\mbox{--}800~{\rm nm}$ light. This is approximately $2\mbox{--}4$ times larger than the $\sigma_{\rm STED}$ reported for NV centers and nearly as large as that of organic fluorophores commonly used in STED microscopy \cite{KAS2004,RIT2007}. We demonstrate STED microscopy on isolated SiV centers in diamond, realizing a resolution $\Delta d=89\pm2~{\rm nm}$, limited by the available STED laser pulse energy ($0.4~{\rm nJ}$). If these properties are similar in sub-10-nm nanodiamonds, and higher STED pulse energies are available, SiV centers may be ideal probes for high resolution STED microscopy in biological systems. Our methods can also be applied to resolving nanoscale SiV center arrays in quantum information applications \cite{TAM2014,SCH2017}.

\section{Experimental Setup}
\label{sec:Experimental Setup}
The SiV optical transitions and emission spectrum are shown in Figs.~\ref{fig:experimental setup}a and~\ref{fig:experimental setup}b, respectively. The pulse sequence used for STED microscopy is shown in Fig.~\ref{fig:experimental setup}c. A laser pulse ($680\mbox{--}700~{\rm nm}$) excites SiV centers on their absorption phonon sideband. A second pulse ($765\mbox{--}800~{\rm nm}$), with a time-delay of $35~{\rm ps}\lesssim\Delta t\lesssim 100\,{\rm ps}$ (Sec.~\ref{sec:SI Temporal}), stimulates SiV emission on the emission phonon sideband. Fluorescence is collected about the SiV zero-phonon line (ZPL) in the band $733\mbox{--}747~{\rm nm}$. Both excitation and stimulated emission pulses have a temporal full-width-at-half-maximum (FWHM), $\tau_p\approx35\,{\rm ps}$ (Sec.~\ref{sec:SI Temporal}), that is considerably shorter than the SiV excited state lifetime ($\tau_{fl}\approx1.2~{\rm ns}$ \cite{WAN2006}). The sequence is repeated after the laser repetition time, $T_{rep}=12.7~{\rm ns}>>\tau_{fl}$, which is long enough to ensure SiV centers are initialized in their ground state at the start of each sequence. 
A schematic of our SiV STED microscope is shown in Fig.~\ref{fig:experimental setup}d. A supercontinuum source is used to generate both excitation and stimulated emission pulses. The SiV centers studied here were formed from ion implantation and annealing. They were typically ${\sim}50~{\rm nm}$ below the diamond surface with an approximate areal density of $10^6\mbox{--}10^8~{\rm cm^{-2}}$. Section~\ref{sec:SI Sample Prep} contains additional details on the samples and how they were prepared.

\begin{figure*}[htbp]
\centering
\includegraphics[width=0.96\textwidth]{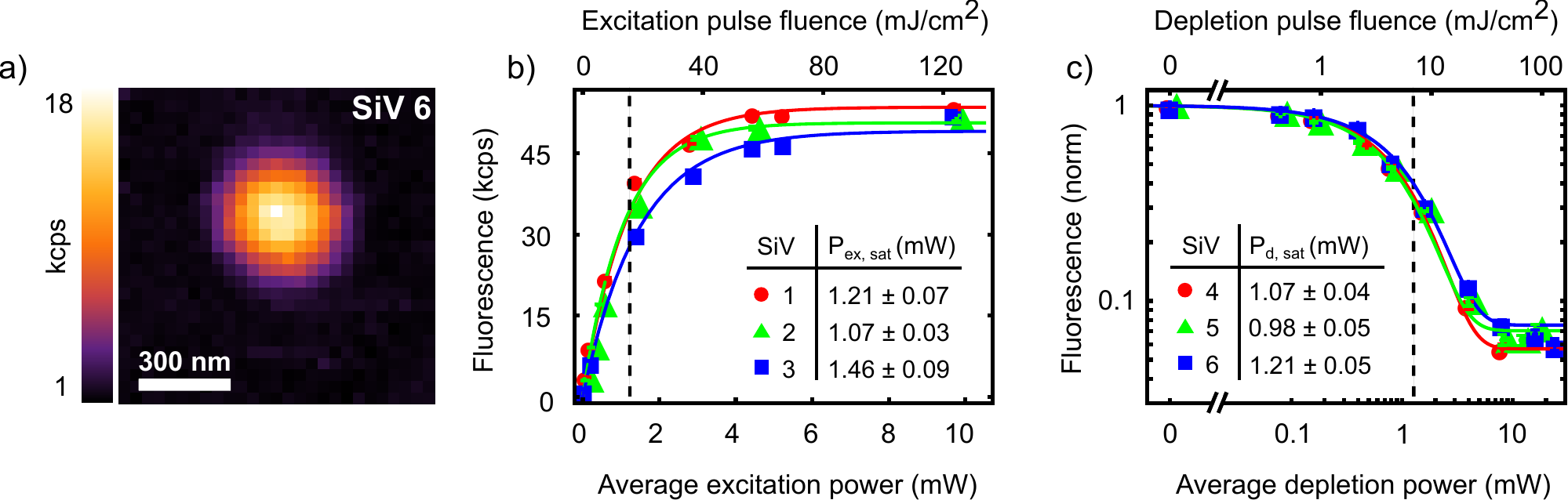}\hfill
\caption{\label{fig:saturation&depletion}
\textbf{Excitation and depletion saturation curves.} (a) Confocal image of ZPL emission ($733\mbox{--}747~{\rm nm}$) from an isolated SiV center excited with $680\mbox{--}700~{\rm nm}$ light. (b) Fluorescence intensity as a function of average excitation power (or corresponding peak pulse fluence) for three SiV centers. Inset: table reporting fitted average excitation saturation powers and fit uncertainties for each SiV center. The mean value is annotated as a dashed line on the plot. (c) Normalized fluorescence intensity as a function of average depletion power (or corresponding peak pulse fluence) for three different SiV centers excited at $P_{\rm ex}\approx P_{\rm ex,\,sat}$. Inset: table reporting fitted average depletion saturation powers and fit uncertainties for each SiV center. The mean value is annotated as a dashed line on the plot.}
\end{figure*}

\section{Results}
Figure~\ref{fig:saturation&depletion}a displays a confocal image of ZPL emission ($733\mbox{--}747~{\rm nm}$) from an isolated SiV center under $680\mbox{--}700~{\rm nm}$ excitation. The FWHM of the feature is ${\sim}270~{\rm nm}$, consistent with the diffraction limit of our microscope. Such isolated features were assumed to be single SiV centers based on their sparsity and nearly identical intensity, Sec.~\ref{sec:SI SiV FL sparsity}. Figure~\ref{fig:saturation&depletion}b shows the detected fluorescence intensity of three SiV centers as a function of average excitation power, $P_{\rm ex}$. We fit these data to a saturation curve of the form $C=C_{\rm max} (1-e^{-P_{\rm ex}/P_{\rm ex,\,sat}})$ \cite{GRE2005}, where $C_{\rm max}$ is the peak detected fluorescence intensity [typically $45$ to $55~{\rm kilocounts/second}$ (kcps) for SiV centers in our setup] and $P_{\rm ex,\,sat}$ is the average excitation saturation power. From the fits, we extract $P_{\rm ex,\,sat}=1.2 \pm 0.2~{\rm mW}$, corresponding to the mean and standard deviation for the set of three SiV centers. By incorporating the laser repetition rate and independently-measured intensity profile of the excitation spot  (Sec.~\ref{sec:SI Cross Section}), this value converts to a saturation pulse fluence $F_{\rm ex,\,sat}= 15 \pm 3~{\rm mJ/cm^2}$. 
The excitation cross-section for this wavelength band is then calculated (Sec.~\ref{sec:SI Cross Section}) as $\sigma_{\rm ex}=E_{\rm ph,\,ex}/F_{\rm ex,\,sat}=(1.8 \pm 0.3)\times10^{-17}~{\rm cm^2}$, where $E_{\rm ph,\,ex}=2.9\times10^{-19}~{\rm J}$ is the excitation photon energy. All remaining experiments were performed with average excitation power $P_{\rm ex}\lesssim P_{\rm ex,sat}$.

We determined the stimulated emission cross section for $765\mbox{--}800~{\rm nm}$ light, $\sigma_{\rm STED}$, using the pulse sequence in Fig.~\ref{fig:experimental setup}c with overlapped Gaussian spatial profiles for excitation and depletion beams. Figure~\ref{fig:saturation&depletion}c shows the normalized fluorescence intensity from three SiV centers as a function of average depletion power, $P_{\rm d}$. These data were fit to an exponential decay function, $C\propto e^{-P_{\rm d}/P_{\rm d,\,sat}}$, revealing an average depletion saturation power $P_{\rm d,\,sat}=1.1\pm0.1~{\rm mW}$ (mean and standard deviation for the three SiV centers). This power corresponds to a depletion saturation pulse fluence $F_{\rm d,\,sat}=6.8 \pm 0.6~{\rm mJ/cm^2}$ (Sec.~\ref{sec:SI Cross Section}). The stimulated emission cross section is therefore $\sigma_{\rm STED}=E_{\rm ph,\,d}/F_{\rm d,\,sat}=(4.0 \pm 0.3)\times10^{-17}~{\rm cm^2}$ (Sec.~\ref{sec:SI Cross Section}), where $E_{\rm ph,\,d}=2.5\times10^{-19}~{\rm J}$ is the depletion photon energy. This cross section is approximately $2\mbox{-}4$ times larger than that of the diamond NV center \cite{HAN2009,RIT2009} and approaches that of the organic dye molecules, ($3\mbox{--}15)\times10^{-17}~\rm cm^2$ \cite{KAS2004,RIT2007}, commonly used in STED microscopy.

\begin{figure*}[ht]
\centering
\includegraphics[width=\textwidth]{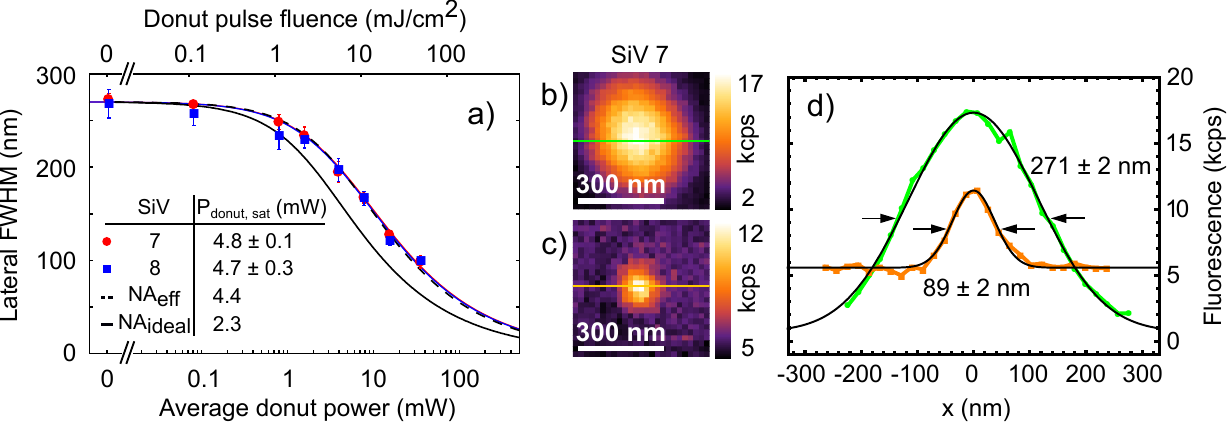}\hfill
\caption{\label{fig:resolution}
\textbf{STED resolution enhancement.} (a) Lateral FWHM of STED profiles of two isolated SiV centers as a function of average donut power (or corresponding peak pulse fluence). Solid red and blue lines are fits to Eq.~\eqref{eqn:STED FWHM} with fitted average donut saturation powers given in the inset. Solid and dashed black lines are the theoretical resolution (Sec.~\ref{sec:SI effective PSF in STED}) for an ideal donut profile ($\rm NA_{ideal}=1.3$) and the experimentally-measured donut profile ($\rm NA_{eff}=1.1$), respectively. A lateral FWHM of ${\sim}20~{\rm nm}$ is expected at $P_{\rm donut}=400~{\rm mW}$, which corresponds to a pulse energy ($5~{\rm nJ}$) commonly used in STED microscopy \cite{WIL2009}. (b) Confocal and (c) STED image of an isolated SiV center taken at $P_{\rm donut}=32~{\rm mW}$ ($0.4~{\rm nJ}$ pulse energy). Annotated linecuts are plotted in (d). The FWHM of the confocal profile (green) is reduced by a factor of three when applying the STED donut beam (orange). Black curves in (d) are Gaussian fits and annotated values are their fitted FWHM.}
\end{figure*}

We next show that STED microscopy applied to SiV centers can be used to realize resolution beyond the optical diffraction limit. We continue to use the pulse sequence in Fig.~\ref{fig:experimental setup}c, but now a vortex phase plate is inserted in the STED path to shape its spatial profile into a donut. We recorded STED images of isolated SiV centers at varying donut powers, $P_{\rm donut}$. Each image is fit to a two-dimensional Gaussian profile to extract the SiV lateral FWHM (Sec.~\ref{sec:SI effective PSF in STED}). At least three images were acquired for each SiV center at each power to determine statistical uncertainty. The results are plotted in Fig.~\ref{fig:resolution}a. Example images taken at $P_{\rm donut}=0$ and $P_{\rm donut}=32~{\rm mW}$ ($0.4~{\rm nJ}$ pulse energy) are shown in Figs.~\ref{fig:resolution}b and c, respectively. The intensity profiles of linecuts through the center of the images are displayed in Fig.~\ref{fig:resolution}d. The FWHM of the confocal image linecut ($P_{\rm donut}=0$) is $271\pm2~{\rm nm}$, consistent with the diffraction-limited resolution of our confocal microscope. At $P_{\rm donut}=32~{\rm mW}$, near the highest power available in our setup, the FWHM shrinks by a factor of ${\sim}3$ to $\Delta d=89\pm2~{\rm nm}$. At this power, we observe a ${\sim}2$-fold reduction in peak fluorescence intensity (see Fig.~\ref{fig:SI Histo}), likely because of imperfect donut contrast. We also observe a slight increase in background due, in part, to anti-Stokes fluorescence (Sec.~\ref{sec:SI Anti-stokes}).

The data in Fig.~\ref{fig:resolution}a were fit to a commonly-used approximation for STED resolution \cite{HAR2008}:
\begin{equation}
\label{eqn:STED FWHM}
\Delta d(P_{\rm donut})\approx\frac{D}{\sqrt{1+\frac{P_{\rm donut}}{P_{\rm donut,\,sat}}}}.
\end{equation}
Here $D$ is the confocal microscope resolution, which we set to $D=270~{\rm nm}$ based on independent measurements, and $P_{\rm donut,~sat}$ is a fitted characteristic power that satisfies $\Delta d(P_{\rm donut,~sat})=D/\sqrt{2}$. From the fits (solid red and blue curves), we extract $P_{\rm donut,~sat}=4.8\pm0.1$ and $4.7\pm0.3~{\rm mW}$ for two different SiV centers. These powers correspond to characteristic peak pulse fluences of $9.7\pm0.2$ and $9.5\pm0.6~{\rm mJ/cm^2}$, respectively (Sec.~\ref{sec:SI effective PSF in STED}).

The theoretical resolution for a perfect donut beam focused with a $\rm NA_{ideal}=1.3$ objective (solid black line in Fig.~\ref{fig:resolution}a) is approximated from a numerical model (Sec.~\ref{sec:SI effective PSF in STED}) incorporating the previously measured $\sigma_{\rm STED}=4\times10^{-17}~{\rm cm^2}$. The corresponding saturation power for this ideal case is $P_{\rm donut,~sat}=2.3~{\rm mW}$, approximately two times smaller than the observed value. Experimentally, we measure a donut beam profile that is more consistent with an effective numerical aperture of $\rm NA_{eff}=1.1$. This may be due to wavefront or polarization distortions of the STED beam and/or under-filling of the beam at the objective's back aperture (see Sec.~\ref{sec:SI effective PSF in STED}). Incorporating this $\rm NA$ into the numerical model (dashed black line in Fig.~\ref{fig:resolution}a), we find excellent agreement with the experimental resolution. The corresponding saturation power, $P_{\rm donut,~sat}=4.4~{\rm mW}$, is consistent with the fits to Eq.~\eqref{eqn:STED FWHM}.

\begin{figure}
\centering
\includegraphics[width=0.8\columnwidth]{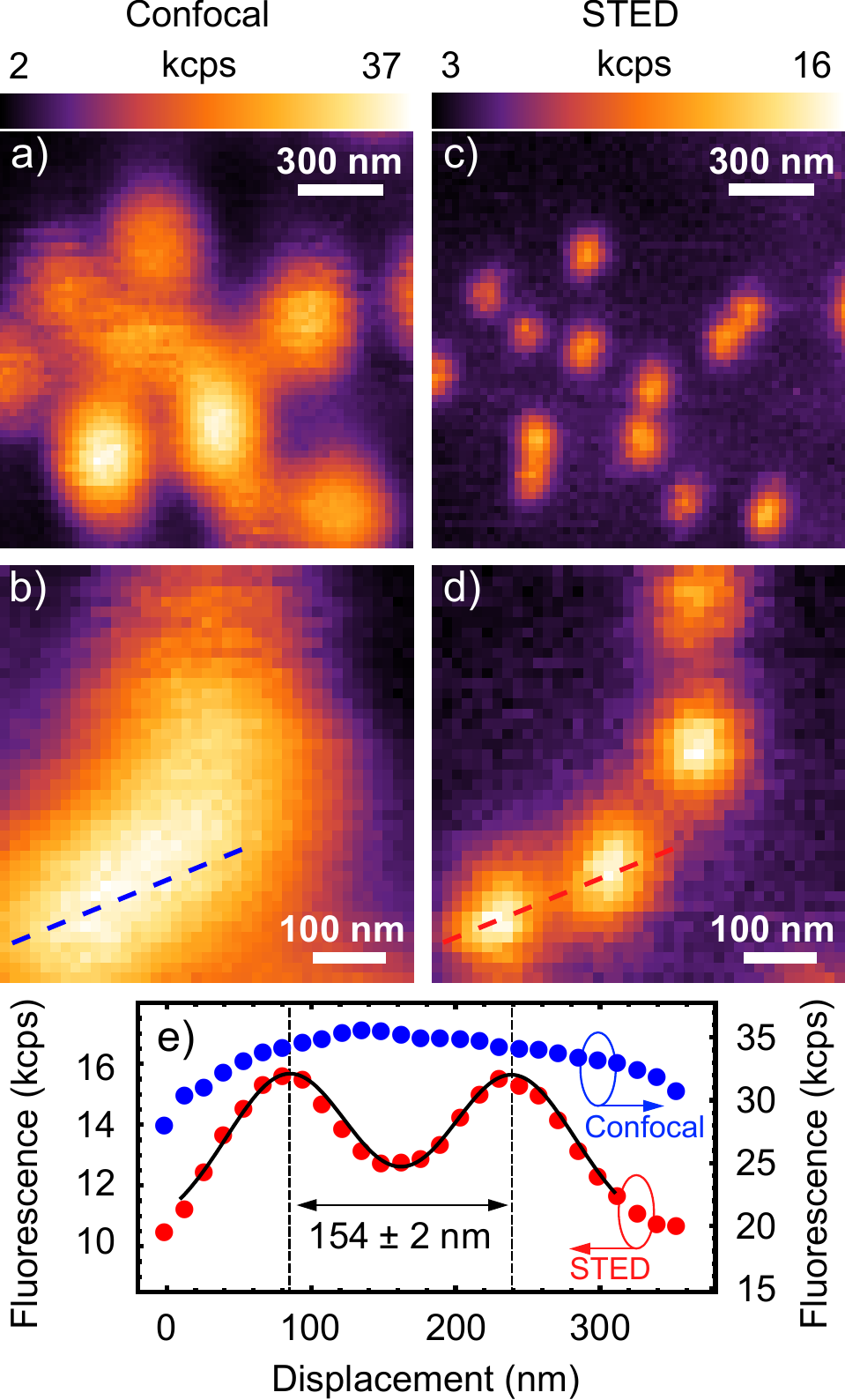}\hfill
\caption{\label{fig:dense SiV}
\textbf{Resolving SiV Clusters in Diamond.} (a,b) Confocal and (c,d) corresponding STED images ($P_{\rm donut}=32~{\rm mW}$) of SiV clusters in two different high-SiV-density regions. The pixel dwell time was $0.1~{\rm s}$. For (a,c) the total image acquisition time was $6~{\rm minutes}$, and for (b,d) it was $3~{\rm minutes}$. (e) Linecuts of the confocal (blue) and STED (red) images across the dashed lines annotated in (b) and (d), respectively. The black solid line is a fit to two Gaussian functions, revealing a SiV center separation of $154\pm2~{\rm nm}$.}
\end{figure}

Finally, we used STED microscopy to resolve SiV centers spaced closer than the optical diffraction limit. Figure~\ref{fig:dense SiV} compares confocal and STED images of SiV clusters in two different high-SiV-density regions (Sec.~\ref{sec:SI Sample Prep}). Unlike the confocal images (Figs.~\ref{fig:dense SiV}a,b), the STED images (Fig.~\ref{fig:dense SiV}c,d) clearly resolve SiV centers separated by ${\lesssim}150~{\rm nm}$. Taking into account the similar brightness and FWHM of features in the STED images (see Sec.~\ref{sec:SI SiV FL sparsity}), it is likely that each individual SiV center in the scan region is resolved. Figure~\ref{fig:dense SiV}e shows linecuts through a sub-region containing closely-spaced SiV centers (dashed lines in Fig.~\ref{fig:dense SiV}b,d). While the confocal image contains little information about the SiV locations, Gaussian fits to the STED linecut reveal two SiV centers separated by $154\pm2~{\rm nm}$.

\section{Discussion and conclusion}
The demonstration of super-resolution STED microscopy with SiV centers has implications for several applications. Importantly, all SiV centers studied here showed perfect photostability (no blinking or bleaching), even under continuous illumination with high STED intensity for several days. However, future work is needed to validate the utility of SiV STED microscopy in biological samples. The modest resolution realized here (${\sim}90~{\rm nm}$) was limited by the maximum STED pulse energy ($\sim 0.4~{\rm nJ}$) available in our setup. If a realistic pulse energy of $5~\rm{nJ}$ was used, the resolution would improve to $\Delta d\approx20~{\rm nm}$ for an optimized STED beam profile (Fig.~\ref{fig:resolution}a). This compares favorably to the STED resolution realized with organic dye molecules ($\Delta d\approx35~{\rm nm}$) under similar conditions \cite{WIL2008, WIL2009}. 

Widespread adoption of SiV probes in STED microscopy will also require development of high-yield methods for fabricating monodisperse sub-10-nm SiV-doped nanodiamonds \cite{CRA2019}. If SiV centers in these nanodiamonds have similar photophysical properties as in bulk diamond, as suggested in prior work~\cite{NEU2011,VLA2014,HIG2017}, they may be ideal probes for super-resolution biological imaging. SiV STED microscopy may also be adapted for super-resolution thermal imaging~\cite{NEU2013,NGU2018} or multiphoton microscopy~\cite{HIG2017}. In addition, our microscope is well suited for the study of nanoscale arrays of SiV centers for applications in quantum information~\cite{TAM2014,SCH2017}. 

In summary, we demonstrated that SiV centers can be used as photostable fluorophores in STED microscopy. We determined the SiV stimulated-emission cross section for $765\mbox{--}800~{\rm nm}$ light to be $\sigma_{\rm STED}=(4.0\pm0.3){\times}10^{-17}~{\rm cm^2}$, a factor of $2\mbox{--}4$ larger than that of NV centers and approaching that of common organic dye molecules. Our results hold promise for future applications in biological imaging and quantum information. 
\\
\\
\begin{acknowledgments}
We gratefully acknowledge advice and support from A.~Laraoui, I.~Fescenko, J.-C. Diels, A.~Rastegari, N.~Mosavian, J.~Damron, N.~Ristoff, M.~D.~Aiello, P.~Kehayias, A.~Jarmola, A.~S.~Backer, P.~R.~Hemmer, K.~A.~Lidke, and C.~Oncebay. We acknowledge the use of nanofabrication and characterization resources at the Department of Energy Center for Integrated Nanotechnologies (CINT).\\
\textbf{Competing interests.} The authors declare no competing financial interests.\\
\textbf{Author contributions.} All authors contributed to the conception and design of the experiment, collection and analysis of the data, and writing of the manuscript. \\
\textbf{Funding.} This work was supported by a Beckman Young Investigator Award and National Science Foundation grant (DMR 1809800).

\end{acknowledgments}
\appendix

\bibliographystyle{apsrev4-1}
\widetext
\clearpage
\begin{center}
\textbf{\large Supplemental Information}
\end{center}
\setcounter{equation}{0}
\setcounter{section}{0}
\setcounter{figure}{0}
\setcounter{table}{0}
\setcounter{page}{1}
\setcounter{equation}{0}
\setcounter{figure}{0}
\setcounter{table}{0}
\setcounter{page}{1}
\makeatletter

\renewcommand{\thetable}{S\arabic{table}}
\renewcommand{\theequation}{S\Roman{section}-\arabic{equation}}

\renewcommand{\thefigure}{S\arabic{figure}}
\renewcommand{\thesection}{S\Roman{section}}
\renewcommand{\bibnumfmt}[1]{[S#1]}
\renewcommand{\citenumfont}[1]{S#1}

\section{Microscope setup}
\label{sec:SI Setup}
A detailed diagram of the STED microscope is shown in Fig.~\ref{fig:experimental setup}d in the main text. Here we provide additional details. A supercontinuum fiber laser (SuperK EXTREME EXR-20, NKT Photonics) provides a train of picosecond optical pulses with a repetition rate ($1/T_{rep}=78~{\rm MHz}$). A polarizing beamsplitter (PBS202, Thorlabs) splits the supercontinuum light into two paths (one for excitation, the other for STED) with orthogonal linear polarizations. Spectral filters are used to select the desired excitation and STED wavelength bands. For excitation ($680\mbox{--}700~{\rm nm}$), a combination of a band-pass filter (FB700-40, Thorlabs) and short-pass filter (FES0700, Thorlabs) are used. For the STED path ($765\mbox{--}800~{\rm nm}$), a combination of a tunable long-pass filter (TL01-290-25x36, Semrock) and short-pass filter (FES0800, Thorlabs) are used. Both beams are expanded and collimated to fill the back aperture ($\sim6~{\rm mm}$ diameter) of an oil-immersion microscope objective (UPLFLN 100x /1.3NA, Olympus) which has $\sim80\%$ transmission for $680\mbox{--}800~{\rm nm}$ light. Dichroic mirrors DM2 (T720lpxr, Chroma) and DM1 (FF765-Di01-25x36x2.0, Semrock) are used to re-combine the excitation and STED beams and reflect away the ZPL emission, as indicated in Fig.~\ref{fig:experimental setup}d. For STED microscopy, a $0\mbox{--}2\pi$ vortex phase plate (VPP-1b, RPC Photonics) is placed in the STED path to generate a donut-shaped intensity profile. A quarter-wave plate (WPQ10ME-780, Thorlabs) placed immediately before the objective lens ensures that the STED beam is right-hand circularly polarized. This polarization preserves the azimuthal symmetry of the donut beam under high-NA focusing \cite{HAO2010}.

Sample fluorescence was collected by the same objective lens, reflected to the emission path by DM1, and focused by a $200\mbox{-}{\rm mm}$ focal length tube lens (ITL200, Thorlabs) onto a $75\mbox{-}{\rm \mu m}$-diameter pinhole (P75H, Thorlabs). The diameter of the pinhole was selected to be approximately equal to the diameter of the ZPL emission Airy disc in the pinhole image plane. Light exiting the pinhole was re-collimated with a lens and passed through a $740\pm6.5~{\rm nm}$ bandpass filter (FF01-740/13, Semrock) to isolate SiV ZPL emission ($733\mbox{--}747~{\rm nm}$). The light was then focused by another lens into a multi-mode fiber (M31L01, Thorlabs) and detected by an avalanche photodiode (SPCM-AQRH-13-FC, Excelitas). The detector output was connected to the counter input of a data acquisition card (NI USB-6363, National Instruments). Three-dimensional scanning of the sample was achieved by a piezo-nanopositioning stage (TRITOR 101 SG, Piezosystem Jena). To form images, the sample scanning was synchronized with the photon counter via the same data acquisition card. The entire sequence was controlled by a home-built LabVIEW program.

\section{Sample Preparation}
\label{sec:SI Sample Prep}

The two samples used in this study were electronic-grade diamond substrates, grown by chemical vapor deposition, with dimensions $\sim2\times2\times0.5~{\rm mm^3}$. One sample (``ME1'') was newly purchased from Microwave Enterprises and had a manufacturer-specified nitrogen concentration of less than 5 parts per billion. The other sample (``UNM 16'') was repurposed from a previous study~\cite{KEH2017}. This sample had been implanted on both sides with nitrogen ions ($^{15}{\rm N}^{+}$, $8\times10^{13}~{\rm ions/cm^2}$, 200 keV) at a large tilt angle ($\sim 86\degree$). This process resulted in a layer of nitrogen atoms extending from the surface to $\gtrsim50~{\rm nm}$ deep with a density of $\sim2$ parts per million. The sample was subsequently annealed for 4 hours at $800\degree~{\rm C}$ and 2 hours at $1100\degree~{\rm C}$ in a vacuum furnace prior to the re-processing done here. 

Both substrates were cleaned in a tri-acid mixture (1:1:1, nitric:perchloric:sulfuric acids) at $200\degree~{\rm C}$. They were then implanted, at normal incidence, with silicon ions ($^{28}{\rm Si}^{+}$) with a dose of $3\times10^9~{\rm ions/cm^2}$ at an energy of 100 keV, leading to a $\sim50~{\rm nm}$ implantation depth. The implanted samples were then annealed for 4 hours at $800\degree~{\rm C}$ and 2 hours at $1100\degree~{\rm C}$ in a vacuum furnace \cite{KEH2017, EVA2016}. After annealing, ME1 had an areal SiV$^{-}$ density of ${\sim}10^6~{\rm cm^{-2}}$, while UNM16 had a SiV$^{-}$ density of ${\gtrsim}10^8~{\rm cm^{-2}}$. The higher SiV$^{-}$ density in UNM16 is likely due to the presence of nitrogen donors which aid conversion of SiV centers into their negatively-charged state \cite{GAL2013, DHO2018}. Both samples contain NV centers, with UNM 16 having a much higher NV density, but they are not detected under the excitation and emission wavelengths used in this work. When exciting with $\sim1~{\rm mW}$ of $680\mbox{--}700~{\rm nm}$ light and detecting at $733\mbox{--}747~{\rm nm}$, both samples exhibited a relatively low and uniform background of $1\mbox{--}2~{\rm kcps}$ in regions without SiV centers.

We used ME1 for all SiV photophysics and STED resolution experiments shown in Figs.~\ref{fig:saturation&depletion} and \ref{fig:resolution}. We used UNM16 for the STED imaging experiments shown in Fig.~\ref{fig:dense SiV}.

\section{Pulse fluence and cross section calculations}
\label{sec:SI Cross Section}

In order to convert the measured optical power of excitation and depletion beams ($P_{ex}$ and $P_d$, respectively) to a pulse fluence ($F_{ex}$ and $F_d$, respectively), detailed knowledge of the beam profiles in the focal plane is required. For the circular Gaussian profile beams used in Fig.~\ref{fig:saturation&depletion}, the peak pulse fluences are given by:
\begin{equation}
\label{eqn:SI excitation fluence formula}
     F_{ex} = \frac{P_{ex}~T_{\rm rep}}{2\pi c_{ex}^2},
\end{equation}
\begin{equation}
\label{eqn:SI depletion fluence formula}
     F_{d} = \frac{P_{d}~T_{\rm rep}}{2\pi c_{d}^2},
 \end{equation}
where $T_{\rm rep}=12.7~{\rm ns}$ is the laser repetition period, and $c_{ex}$ and $c_d$ are the standard deviations of the Gaussian focal-plane spatial profiles for excitation and depletion beams, respectively.

\begin{figure}[!tb]
\centering
\includegraphics[width=0.6\columnwidth]{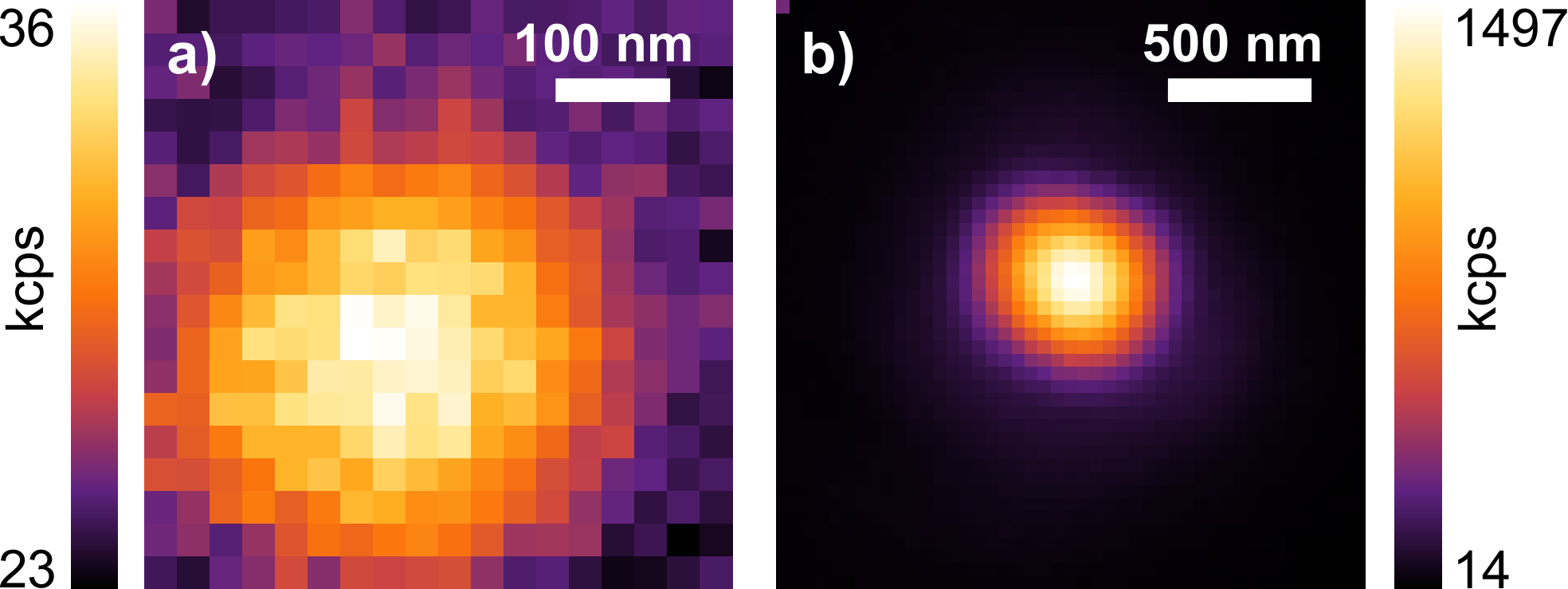}\hfill
\caption{\label{fig:SI SiV&bead area}
\textbf{Excitation and depletion beam profiles.} (a) Scanning confocal image of ZPL emission ($733\mbox{--}747~{\rm nm}$) from an isolated SiV center excited by $680\mbox{--}700~{\rm nm}$ light at 1 ${\rm mW}$ average power. (b) Scanning confocal image of ZPL emission from an individual bead excited by the Gaussian depletion beam ($765\mbox{--}800~{\rm nm}$) at 20 ${\rm \micro W}$ average power. The pinhole in the emission path was removed during the scans in both (a) and (b) to make an accurate measurement of the beam profile.}
\end{figure}

To determine $c_{ex}$, scanning confocal fluorescent images of isolated SiV centers were recorded, Fig.~\ref{fig:SI SiV&bead area}a. Here the pinhole was removed from the emission path to faithfully image the beam profile. The SiV centers were excited by $680\mbox{--}700~{\rm nm}$ light at a power below saturation. Several images were recorded and fit to circular Gaussian profiles, revealing $c_{ex}=127\pm2~{\rm nm}$. 

To determine $c_{d}$, scanning confocal fluorescent anti-Stokes images (again with pinhole removed) of individual fluorescent beads (Infrared fluorescent 715/755, $0.1~{\rm \mu m}~{\rm FluoSpheres}$, ThermoFisher Scientific F8799) were recorded, Fig.~\ref{fig:SI SiV&bead area}b. The beads were diluted and spread on a cover-slip, then excited by the Gaussian depletion beam ($765\mbox{--}800~{\rm nm}$) at low power ($20~{\rm \upmu W}$). Gaussian fits to several bead images revealed $c_{d}=182\pm8~{\rm nm}$.

The one-photon absorption cross sections for excitation and stimulated emission are defined as:
\begin{equation}
\label{eqn:SI excitation Cross Section 1}
     \sigma_{\rm ex} = \frac{E_{\rm ph,\,ex}}{F_{\rm ex,\,sat}},
 \end{equation}
 \begin{equation}
 \label{eqn:SI depletion Cross Section 1}
     \sigma_{\rm STED} = \frac{E_{\rm ph,\,d}}{F_{\rm d,\,sat}},
 \end{equation}
 where $E_{\rm ph,\,ex}$ and $E_{\rm ph,\,d}$ are the excitation and depletion photon energies, $F_{\rm ex,\,sat}$ is the saturation peak pulse fluence of the excitation beam (corresponding to relative excited-state population of $1-1/e$), and $F_{\rm d,\,sat}$ is the saturation peak pulse fluence of the depletion beam (corresponding to relative excited-state population of $1/e$). 
 
 Using the excitation saturation powers obtained for the three SiV centers shown in Figure~\ref{fig:saturation&depletion}b, three different values for the excitation cross section ($\sigma_{ex}$) were calculated, Eq.~\eqref{eqn:SI excitation Cross Section 1}. The mean value and standard deviation are  $\sigma_{\rm ex}=(1.8 \pm 0.3)\times10^{-17}~{\rm cm^2}$, as reported in the main text. Using the two depletion saturation powers obtained for SiV 4\,\&\,5 shown in Figure~\ref{fig:saturation&depletion}c (we omitted SiV 6 because that data was obtained without measuring the depletion beam's profile using beads immediately beforehand), two values of the stimulated emission cross section ($\sigma_{\rm STED}$) were calculated [Eq.\,\eqref{eqn:SI depletion Cross Section 1}] and we reported their mean value and standard deviation in the text as $\sigma_{\rm STED}=(4.0 \pm 0.3)\times10^{-17}~{\rm cm^2}$.

\section{Temporal characterization of laser pulses}
\label{sec:SI Temporal}
\begin{figure}[!htb]
\centering
\includegraphics[width=0.70\columnwidth]{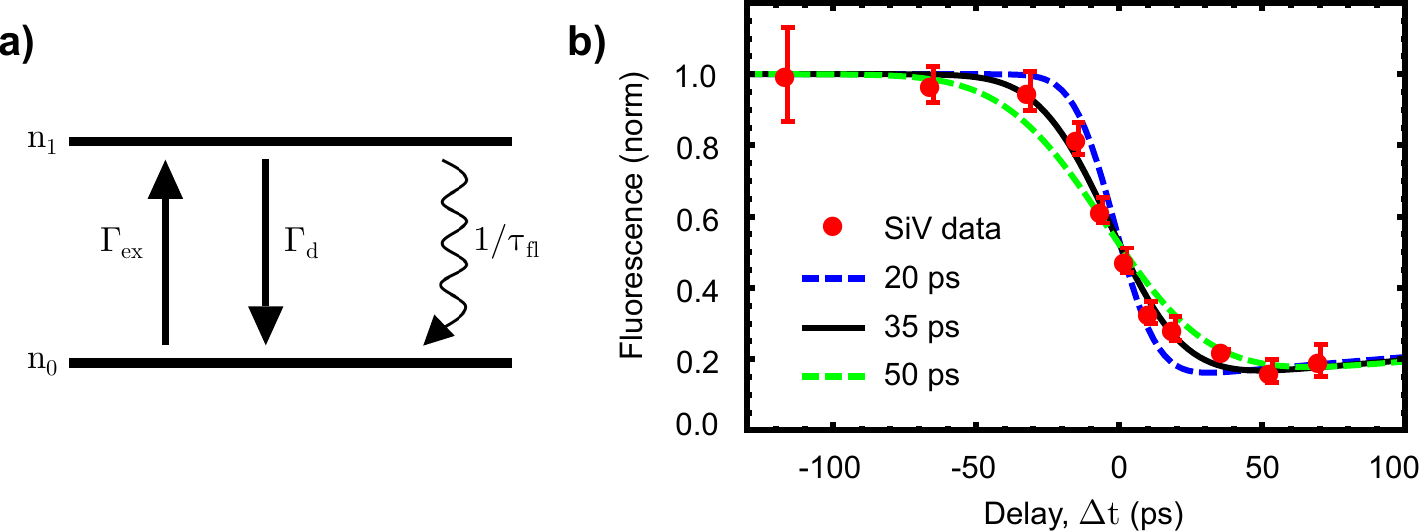}\hfill
\caption{\label{fig:SI Overlap with levels}
\textbf{Depletion efficiency Vs time-delay.} (a) Energy levels of a closed two-level atom. $\Gamma_{\rm ex}$, $\Gamma_{\rm d}$ and $1/\tau_{fl}$ are the excitation, stimulated emission and spontaneous emission rates, respectively. (b) Red circles: normalized fluorescence intensity of an isolated SiV center excited at $P_{ex}=1.5\,{\rm mW}$ and depleted at $P_{d}=5.0\,{\rm mW}$ for different time delays between the excitation and depletion beams. For each delay, three measurements were taken and the data points and error bars are their mean and standard deviation, respectively.  Theoretical curves based on Eq.~\eqref{eqn:SI two-level closed system} are plotted for three FWHM pulse widths, $\tau_p=20,~35,~{\rm and}~50~{\rm ps}$.}
\end{figure}

The temporal properties of the excitation and depletion pulses were determined by monitoring the fluorescence of SiV centers as a function of the delay between excitation and depletion pulses, $\Delta t$. The normalized fluorescence intensity of an isolated SiV center excited ($P_{ex}=1.5\,{\rm mW}$) and depleted ($P_{d}=5.0\,{\rm mW}$) by Gaussian-spatial-profile pulses were obtained as a function of $\Delta t$, shown as red circles in Fig.~\ref{fig:SI Overlap with levels}b.

To describe the dynamics and extract pulse parameters, we model the SiV center as a closed two-level system under non-resonant optical pumping, Fig.~\ref{fig:SI Overlap with levels}a. The excitation and depletion pulses are assumed to have a Gaussian temporal profile. For simplicity, we assume that the FWHM of the temporal profile, $\tau_p$, is the same for both excitation and depletion pulses. Under these assumptions, the time-dependent excited state population of the SiV center, $n_1(t)$, is given by:
\begin{equation}
\label{eqn:SI two-level closed system}
    \frac{d\,n_1 (t)}{dt} = \Gamma_{\rm ex}\,e^{-4\ln{2}\,(\frac{t}{\tau_{p}})^2}[1-n_1 (t)]-[\Gamma_{\rm d}\,e^{-4\ln{2}\,(\frac{t-\Delta t}{\tau_{p}})^2}+\frac{1}{\tau_{fl}}]\,n_1 (t),
\end{equation}
where $\Delta t$ is the time-delay between the excitation and depletion beams, $T_{\rm rep}=12.7~{\rm ns}$ is the pulse sequence repetition period, and $\tau_{fl}=1.2\,{\rm ns}$ is the SiV excited state lifetime. The excitation and stimulated emission rates are defined as $\Gamma_{\rm ex}=F_{\rm ex}/(\tau_p\,F_{\rm ex,\,sat})$ and $\Gamma_{\rm d}=F_{\rm d}/(\tau_p\,F_{\rm d,\,sat})$, respectively. Based on independent measurements of the excitation and depletion powers and pulse shapes used in experiment, we set $F_{\rm ex}/F_{\rm ex,\,sat}=1.25$ and $F_{\rm d}/F_{\rm d,\,sat}=2.0$.

To model the fluorescence intensity, solutions to Eq.~\eqref{eqn:SI two-level closed system} are obtained numerically and the excited-state population is integrated from $t=0$ to $t=T_{\rm rep}$. 
We assume that at the beginning of each sequence the SiV center is in the ground state, $n_1 (0)=0$. In Fig.~\ref{fig:SI Overlap with levels}b, the normalized integrated excited-state population is plotted as a function of time delay ($\Delta t$) for three different values of pulse width ($\tau_p$). The FWHM pulse length that best matches the experimental data is $\tau_p=35~{\rm ps}$. 

It can be seen from Fig.~\ref{fig:SI Overlap with levels}b that the depletion efficiency is maximized when $\Delta t\approx \tau_p$. For all experiments reported in the main text, we set the time delay between pulses by maximizing the depletion efficiency. We therefore assume the time delay was in the range $35~{\rm ps}\lesssim\Delta t\lesssim100~{\rm ps}$.

\section{Lateral point-spread function in STED microscopy}
\label{sec:SI effective PSF in STED}

\begin{figure}[b]
\centering
\includegraphics[width=0.7\columnwidth]{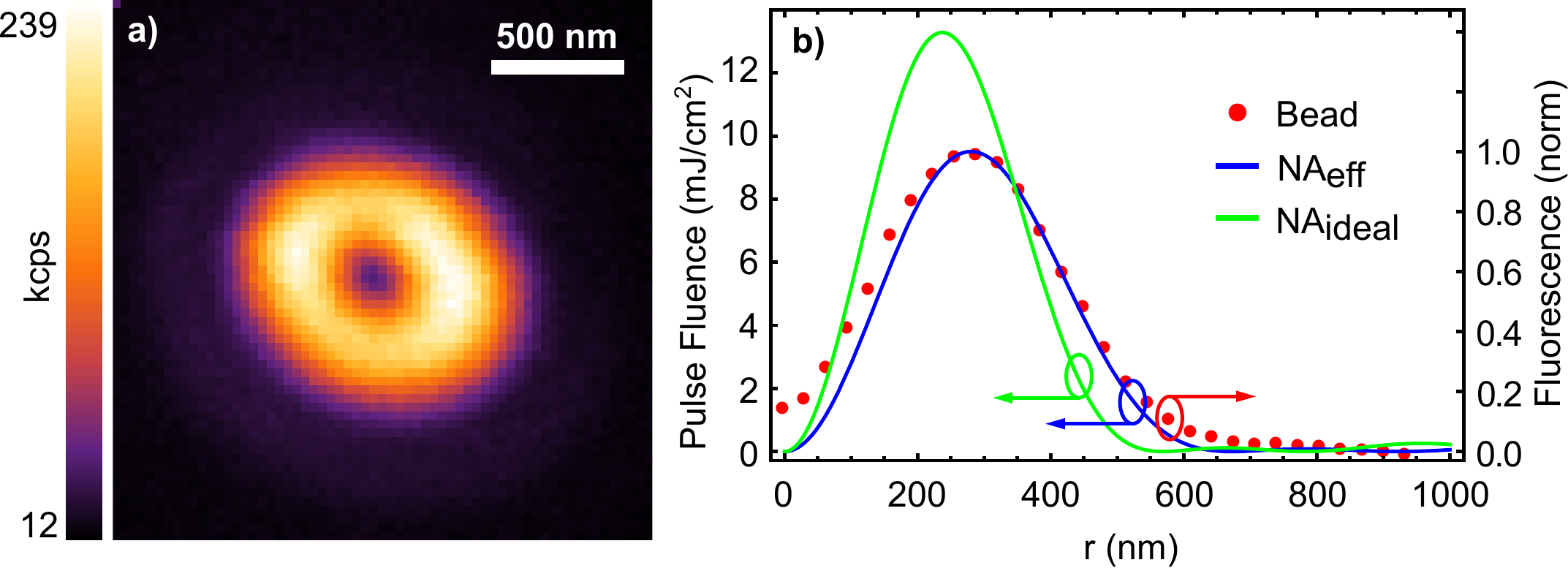}\hfill
\caption{\label{fig:SI Donut}
\textbf{Donut quality.} (a) Scanning confocal image (with pinhole removed) of ZPL emission from an individual bead excited by the donut beam ($765\mbox{--}800~{\rm nm}$). (b) Left axis: theoretical pulse fluence profile of a donut-shaped beam, Eq.~\eqref{eqn:SI Donut Intensity}, as a function of radial distance. The profile is plotted for $P_{\rm donut}=4.75\,{\rm mW}$ and $\lambda=790\,{\rm nm}$ with $\rm NA_{eff}=1.1$ (blue) and $\rm NA_{ideal}=1.3$ (green). Right axis: the normalized fluorescence intensity profile of the bead image in (a) determined from the average of four line-cuts beginning at the donut's center.}
\end{figure}

The donut quality of the STED beam plays a major role in achieving high resolution in STED microscopy. To measure the experimental donut profile, we recorded a scanning confocal (but with pinhole removed) anti-Stokes fluorescent image of an individual bead excited by our donut-shaped STED beam, Figure~\ref{fig:SI Donut}a. An average of four line-cuts beginning at the donut' s center is shown as red circles in Fig.~\ref{fig:SI Donut}b.

To compare to the theoretical optimal donut profile, we assume that the vortex phase plate converts a coherent plane wave into an ideal Laguerre-Gaussian donut beam. The donut pulse fluence profile in the focal plane of the objective lens, $F_{\rm donut}(r)$, can then be approximated as \cite{KHO1992}:

\begin{equation}
\label{eqn:SI Donut Intensity}
    F_{\rm donut}(r) =P_{\rm donut}\,T_{\rm rep}\frac{\pi \rm{NA}^2}{\lambda^2} \left(\pi \frac{H_0(u)J_1(u)-H_1(u)J_0(u)}{u}\right)^2.
\end{equation}
Here $P_{\rm donut}$ is the average donut power, $T_{\rm rep}=12.7~{\rm ns}$ is the pulse repetition period, $u=2\pi r {\rm NA}/\lambda$ is the normalized radial distance, $\lambda$ is the wavelength of the donut beam, $\rm NA$ is the objective numerical aperture, $J_0$ and $J_1$ are the zeroth and first order Bessel Functions, and $H_0$ and $H_1$ are the zeroth and first order Struve Functions. 

Figure~\ref{fig:SI Donut}b shows plots of $F_{\rm donut}(r)$, calculated from Eq.~\eqref{eqn:SI Donut Intensity}, for two values of $\rm NA$. The two values correspond to the true objective numerical aperture, $\rm NA_{ideal}=1.3$, and an effective numerical aperture, $\rm NA_{eff}=1.1$, that best fits the experimental profile. The difference between the optimal donut profile ($\rm NA_{ideal}=1.3$) and the experimental profile ($\rm NA_{eff}=1.1$) may be due to a combination of imperfect circular polarization, deviations from the plane wave approximation before the vortex phase plate, and/or under-filling of the beam at the objective's back aperture. 

The $\rm NA_{eff}=1.1$ profile was used to convert the average donut power, $P_{\rm donut}$, into peak donut pulse fluence in Fig.~\ref{fig:resolution}a. The characteristic saturation powers, $P_{\rm donut,\,sat}=4.7$ and $4.8\,{\rm mW}$, obtained for the two isolated SiV centers shown in Fig.~\ref{fig:resolution}a correspond to peak donut pulse fluences of $9.5$ and $9.7~{\rm mJ/cm^2}$, respectively. 

\begin{figure}[H]
\centering
\includegraphics[width=0.40\columnwidth]{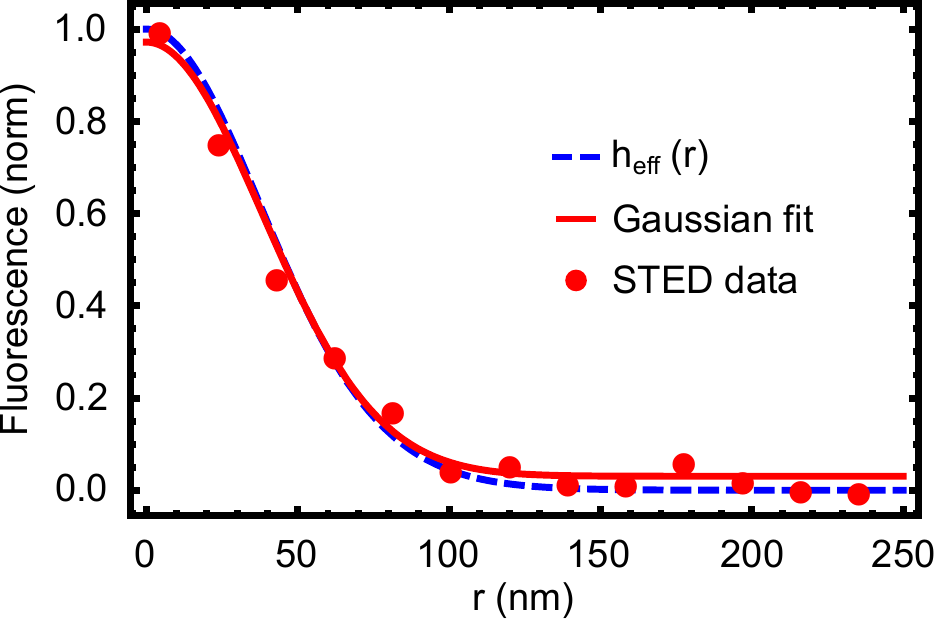}\hfill
\caption{\label{fig:Gaussian fit Vs Narrowing function}
\textbf{The effective PSF in STED microscopy.} Red circles: normalized STED fluorescence intensity profile of SiV 7 using $P_{\rm donut}=32\,{\rm mW}$ (Fig.~\ref{fig:resolution}d). Solid red line: a Gaussian fit to the data revealing a FWHM $\Delta d\approx90~{\rm nm}$. Dashed blue line: theoretical STED PSF, $h_{eff}(r)$, determined from Eqs.~\eqref{eqn:SI Donut Intensity} and \eqref{eqn:SI STED PSF}. The parameters used in the calculation are: $D=270\,{\rm nm}$, $P_{\rm donut}=32\,{\rm mW}$, $\lambda=790\,{\rm nm}$, $\rm NA_{eff}=1.1$ and $F_{\rm d,\,sat}=6.8~{\rm mJ/cm^2}$.}
\end{figure}

To understand the relationship between donut quality and STED resolution, we define the lateral STED point-spread function (PSF) as \cite{HAR2008S}:
\begin{equation}
\label{eqn:SI STED PSF}
    h_{eff}(r) = h_c(r)~e^{-F_{\rm donut}(r)/F_{\rm d, sat}},
\end{equation}
where $h_c(r)\approx e^{-4\ln{2}\,(r/D)^2}$ is the confocal PSF with a FWHM of $D=270~{\rm nm}$. Figure~\ref{fig:Gaussian fit Vs Narrowing function} plots $h_{eff}(r)$ using $F_{\rm d,\,sat}=6.8~{\rm mJ/cm^2}$. The pulse fluence, $F_{\rm donut}(r)$ was computed from Eq.~\eqref{eqn:SI Donut Intensity} using $\rm NA_{eff}=1.1$, $\lambda=790~{\rm nm}$, and $P_{\rm donut}=32~{\rm mW}$ (the highest power used in our experiments). The theoretical STED PSF is in excellent agreement with the experimental PSF obtained at this power and is well approximated by a circular Gaussian function with FWHM $\Delta d=90~{\rm nm}$. For all experiments in the main text, we used the circular Gaussian profile approximation to extract the resolution to simplify analysis.

\section{Fluorescence intensity distribution of isolated SiV centers}
\label{sec:SI SiV FL sparsity}
\begin{figure}[H]
\centering
\includegraphics[width=0.6\columnwidth]{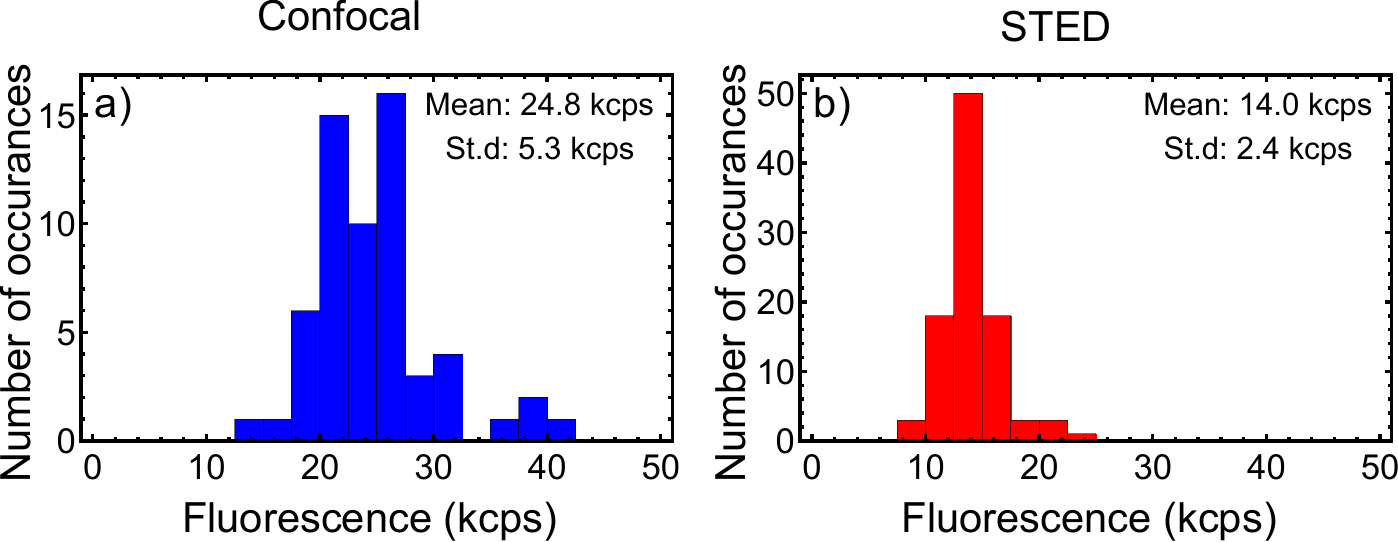}\hfill
\caption{\label{fig:SI Histo}
\textbf{Fluorescence intensity distribution of isolated SiV centers.} Distribution of fluorescence intensity of isolated SiV centers in (a) confocal and (b) STED images. Data were obtained from the same region on the high-density sample (SiV density: ${\sim}10^8~{\rm cm^{-2}}$) within an area of $10\times10~{\rm \upmu m^2}$ at $P_{ex}=2.1~{\rm mW}$ and $P_{\rm donut}=32~{\rm mW}$. The insets in (a) and (b) are the mean number and standard deviation of the corresponding distributions.}
\end{figure}

In order to determine whether the isolated SiV centers in our samples are really single emitters or not, we recorded large fluorescent images ($10\times10\, {\rm \mu m}^2$) of the dense sample in both confocal and STED configurations. The distribution of the peak SiV fluorescence intensities in both cases is shown in Fig.~\ref{fig:SI Histo}. The narrow distribution suggests that the features likely arise from single emitters with relatively homogenous photophysical properties. However, the main conclusions in this work (STED cross section, resolution, etc.) would remain valid even if these isolated fluorescent features came from multiple emitters.

\section{Anti-Stokes excitation}
\label{sec:SI Anti-stokes}
As discussed in the main text, a faint halo background can sometimes be observed in STED images of isolated SiV centers. This background follows closely the STED donut profile and likely arises from anti-Stokes emission. At room temperature, the SiV center has a small (but non-zero) probability of being in an excited vibrational level within the ground-state manifold \cite{TRA2019}. Thus the STED beam has a small probability to excite SiV centers in addition to its primary role of stimulating emission from the excited state. For high STED intensities, this anti-Stokes excitation phenomenon can reduce the contrast of STED images and limit the achievable resolution  \cite{HEL2007,VIC2012}. Thus, for STED microscopy with fluorophores having a relatively small Stokes shift, as is the case for SiV centers (Fig.~\ref{fig:experimental setup}b), there is a trade off between increasing $\sigma_{\rm STED}$ (by exciting at the peak of the phonon sideband) and introducing background due to anti-Stokes excitation.

\begin{figure}[H]
\centering
\includegraphics[width=0.65\columnwidth]{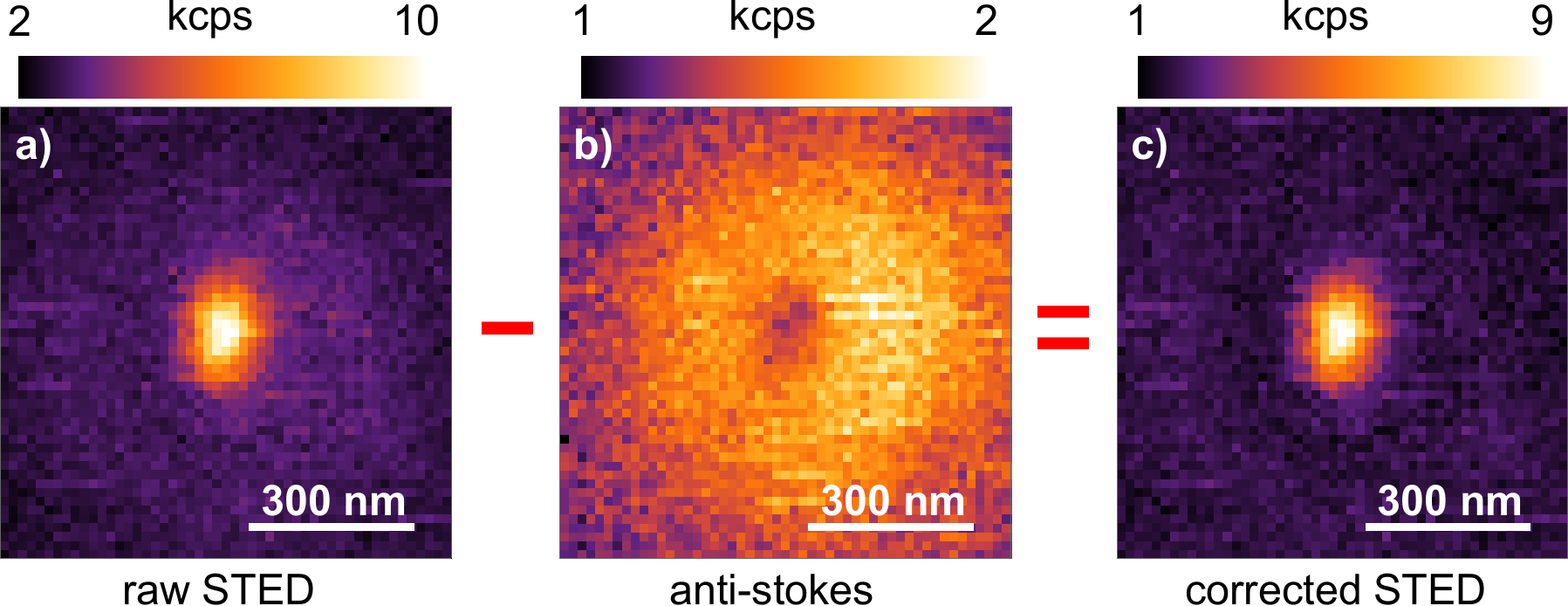}\hfill
\caption{\label{fig:SI Anti-Stokes SiV donut 1}
\textbf{Anti-Stokes excitation of a SiV center.} (a) Raw STED image of an isolated SiV center taken at $P_{\rm ex}=1.1~{\rm mW}$ and $P_{\rm donut}=35.2~{\rm mW}$. (b) Weighted anti-Stokes fluorescent image of the same SiV center taken at $P_{\rm ex}=0~{\rm mW}$ and $P_{\rm donut}=35.2~{\rm mW}$. (c) The corrected STED image after subtracting the anti-Stokes background fluorescence.}
\end{figure}

Figure~\ref{fig:SI Anti-Stokes SiV donut 1}a shows a STED image of an isolated SiV center taken at the highest available donut power in our setup, ($P_{\rm donut}=35.2\,{\rm mW}$). A weak halo of background fluorescence is observed. By blocking the excitation beam and recording another image, Fig.\,\ref{fig:SI Anti-Stokes SiV donut 1}b, it is seen that this weak background ($\sim 1~{\rm kcps}$) follows the shape of the donut profile. This background image can be subtracted from the raw STED images in order to improve the image contrast, Fig.~\ref{fig:SI Anti-Stokes SiV donut 1}c.

\bibliographystyle{apsrev4-1}

\end{document}